\documentstyle[12pt]{article}

\def\be{\begin{equation}}
\def\ee{\end{equation}}
\def\bea{\begin{eqnarray}}
\def\eea{\end{eqnarray}}

\def\GeV{\,{\rm GeV}}

\def\keV{\,{\rm keV}}
\def\MeV{\,{\rm MeV}}
\def\sec{\,{\rm sec}}
\def\Gyr{\,{\rm Gyr}}
\def\yr{\,{\rm yr}}
\def\rcm{\,{\rm cm}}

\def\Mpc{\,{\rm Mpc}}

\def\eV{{\,\rm eV}}

\def\cmm2{{\,\rm cm^{-2}}}
\def\cm2{{\,{\rm cm}^2}}
\def\cmm3{{\,{\rm cm}^{-3}}}
\def\gcmm3{{\,{\rm g\,cm^{-3}}}}
\def\kms{\,{\rm km\,s^{-1}}}

\def\la{\mathrel{\mathpalette\fun <}}
\def\ga{\mathrel{\mathpalette\fun >}}
\def\fun#1#2{\lower3.6pt\vbox{\baselineskip0pt\lineskip.9pt
  \ialign{$\mathsurround=0pt#1\hfil##\hfil$\crcr#2\crcr\sim\crcr}}}

\begin{document}
\begin{flushright}

FERMILAB-Conf-95/034-A \\
astro-ph/9503017

\end{flushright}

\begin{center}
{\Large \bf The Hot Big Bang and Beyond\footnote{To be published
in the Proceedings of CAM-94 (Cancun, Mexico, September 1994)
and the Proceedings of Diffuse Background Radiation IAU 168
(The Hague, August 1994)}}
\end{center}

\begin{center}
{{\large \bf Michael S. Turner} \\[1cm]

{\it Departments of Physics and of Astronomy \& Astrophysics\\
Enrico Fermi Institute, The University of Chicago,
Chicago, IL~~60637-1433}\\

\vspace{.1in}

{\it NASA/Fermilab Astrophysics Center\\
Fermi National Accelerator Laboratory, Batavia, IL~~60510-0500}}
\end{center}
\noindent
\begin{center}
\begin{small}
\begin{minipage}[t]{12cm}
{\bf Abstract.}
The hot big-bang cosmology provides a reliable accounting
of the Universe from about $10^{-2}\sec$ after the bang until the
present, as well as a robust framework for speculating back to
times as early as $10^{-43}\sec$.  Cosmology faces a number of important
challenges; foremost among them are determining the quantity and composition of
matter in the Universe and developing a detailed and coherent picture
of how structure (galaxies, clusters of galaxies, superclusters,
voids, great walls, and so on) developed.  At present there is a working
hypothesis---cold dark matter---which is based upon inflation
and which, if correct, would extend the big bang model
back to $10^{-32}\sec$ and cast important light on the unification
of the forces.  Many experiments and observations, from
CBR anisotropy experiments to Hubble Space Telescope observations
to experiments at Fermilab and CERN,
are now putting the cold dark matter theory to the test.
At present it appears that the theory is viable only if the
Hubble constant is smaller than
current measurements indicate (around $30\kms\Mpc^{-1}$),
or if the theory is modified slightly, e.g.,
by the addition of a cosmological constant,
a small admixture of hot dark matter ($5\eV$ ``worth
of neutrinos''), more relativistic particles, or a tilted
spectrum of density perturbations.

\end{minipage}
\end{small}
\end{center}
\setlength{\baselineskip}{1\baselineskip}

\section{Successes}

The success of the hot big-bang cosmology (or standard cosmology
as it is known) is simple to describe:  It provides a reliable and
tested accounting of the Universe from a fraction of a second
after the bang (temperatures of order a few MeV) until the
present 15 Billion years later (temperature 2.726 K).  When
supplemented by the standard model of particle
physics and various ideas about physics at higher energies
(e.g., supersymmetry, grand unification, and superstrings)
it provides a sound foundation for speculations
about the Universe back to $10^{-43}\sec$
after the bang (temperatures of $10^{19}\GeV$) and perhaps even earlier
\cite{cosmology}.

The fundamental observational data that support the standard
cosmology are:  the universal expansion (Hubble flow of galaxies);
the cosmic background radiation (CBR);
and the abundance of the light elements D, $^3$He, $^4$He, and
$^7$Li.  The Hubble law ($z\simeq v/c \simeq
H_0 d$) has been tested to a redshift $z\sim 0.05$ \cite{mould} and
the highest redshift object is a QSO with $z=4.90$.  (One plus redshift
is the size of the Universe today
relative to its size at the time of emission, $1+z = R_0/R_E$;
R is the cosmic scale factor, or relative size of the Universe).

The surface of last scattering for the CBR is the Universe at
an age of a few hundred thousand years ($T\sim 0.3\eV$ and
redshift $z\sim 1100$).  COBE has determined its temperature to
be $2.726\pm 0.005\,$K and constrains any deviations from a
black-body spectrum to be less than 0.03\% \cite{firas}.
The CBR temperature is very uniform:
the difference between two points separated by angles from
arcminutes to $90^\circ$ is less than $300\mu K$, indicating
that the Universe had a very smooth beginning.  There is a dipole
anisotropy in the CBR temperature of about $3\,$mK,
due to our motion with respect to the cosmic rest frame
(the ``peculiar velocity'' of the Local Group is
$620\kms$ toward the constellation
Leo), and temperature differences on angular
scales from $0.5^\circ$ to $90^\circ$ have been detected
by about ten experiments at the level of about $30\mu$K \cite{wss}.

The abundance of the light elements, which range from about
24\% for $^4$He to $10^{-5}$ for D and $^3$He and $10^{-10}$
for $^7$Li are consistent with the predictions of big-bang nucleosynthesis.
The synthesis of the light elements occurred when the Universe
was of order seconds old and the temperature was of order MeV.
Big-bang nucleosynthesis is the earliest test of the standard cosmology,
and it passes it with flying colors \cite{copi}.

Finally, the standard cosmology provides a general
framework for understanding
how the Universe evolved from a very smooth beginning to the
abundance of structure observed today---galaxies, clusters of galaxies,
superclusters, voids, great walls and so on.  Small (primeval)
variations in the matter density ($\delta\rho /\rho \sim 10^{-5}$)
were amplified by gravity (the Jeans' instability in the expanding
Universe) eventually resulting in the structure seen today \cite{peebles}.
The CBR temperature fluctuations detected
on angular scales from $0.5^\circ$ to $90^\circ$ are strong evidence
for the existence of primeval density fluctuations.

In addition to accounting for the evolution of
the Universe from 0.01 sec onward,
the standard cosmology provides a sound framework
for speculating about even earlier times---at least back to the Planck
time ($10^{-43}\sec$ and temperature of order $10^{19}\GeV$).
Of course, advances in particle theory
have played an important role here.

According to the standard model of particle physics
the fundamental particles are point-like quarks and leptons
whose interactions are weak enough to treat perturbatively.
The cosmological implications of this are profound:
The Universe at temperatures
greater than about $150\MeV$ (times earlier than $10^{-5}\sec$)
consisted of a hot, dilute gas of quarks, leptons, and gauge bosons
(photons, gluons, and at high enough temperatures $W$ and
$Z$ bosons, the carriers of the electromagnetic, strong and weak forces).
The standard model of particle physics, which has been tested up to
energies of several hundred GeV, provides the microphysics
needed to discuss times as early as $10^{-11}\sec$.  In addition,
it provides a firm platform for speculations about the
unification of forces and particles
(e.g., supersymmetry and grand unification), and in
turn, the necessary microphysics
for extending cosmological speculations back to
the Planck epoch.  Earlier than the Planck time a quantum description
of gravity is needed, and superstring theory is a good candidate for such.

While the hot big-bang cosmology and modern particle theory
have allowed ``sensible''---and very interesting---speculations
about the early Universe, there is no
evidence {\it yet} that any of these speculations is correct.  However,
contrast this with the situation before the early 1970s.
The count of ``elementary particles'' (baryons and mesons)
had exceeded 100 and was growing exponentially with mass;
this, the strength of their interactions and their finite
sizes precluded any sensible speculation about
the Universe at times earlier than about $10^{-5}\sec$ \cite{weinberg}.

\section{Challenges}

Cosmology is not without its challenges.  In its success, the
hot big-bang model has allowed cosmologists to ask
even deeper questions.  They include:  What is the
quantity and composition of the ubiquitous
dark matter in the Universe?   What is
the origin and nature of the primeval density perturbations that
seeded structure and precisely how did the
structure form?  What is the origin of the
cosmic asymmetry between matter and antimatter?  Why is the
observed portion of the Universe so smooth and flat?  And does
this mean that the entire Universe is the same?  Are there
observational consequences of the phase transitions that
the Universe has undergone (transition from quarks to
nucleons and related particles, electroweak symmetry
breaking, and possibly others) during its earliest moments?
Are there observable consequences of the quantum-gravity epoch?
Why does the Universe have four dimensions?  What caused the expansion
in the first place?

The first two of these challenges, the nature of the
dark matter and the details of structure formation,
are in my opinion the most pressing and may well be resolved soon.
Thus they offer an excellent opportunity for extending the
big-bang cosmology back to much earlier times.

That is not to say that the other challenges are not important
or do not have potential for advancing our understanding.   In addition,
there more practical challenges that I have not listed;
for example, a precise determination of the three traditional
parameters used to describe ``our world model,'' the Hubble constant,
the deceleration parameter, and the cosmological constant,
or an explanation for the primeval magnetic fields required
to seed the magnetic fields seen throughout the Universe today.

\subsection{Discard the big bang?}

There are few who believe the big bang faces challenges of
such enormity that they will led to its downfall \cite{hoyle}.
For example, if the
Hubble constant is as large as some determinations indicate,
say around $80\kms\Mpc^{-1}$ and the oldest stars are as old
as some determinations indicate, say around $16\Gyr$, then
a real dilemma exists because without recourse to a cosmological
constant the time back to the back is less than $12\Gyr$ \cite{hubble}.

Before the COBE discovery of CBR anisotropy in 1992 \cite{dmr},
some argued that the absence of anisotropy
precluded inhomogeneity of a size large enough to seed all
the structure seen today.  The big-bang has weathered that
storm:  Fluctuations in the CBR temperature have been detected
and are now seen on scales from $0.5^\circ$ to $90^\circ$.
In fact, careful calculations indicate that if anything the level
of temperature fluctuations seen is slightly larger than is expected,
given the structure seen today \cite{wss,jpo}.

The only two competitors to the big bang are the quasi steady-state
model \cite{qss} and the plasma universe model.  At the moment
the problems that these models face seem far more daunting:
themalization of starlight to produce 2.726\,K black body
background with no spectral distortion (quasi steady-state)
and the formulation of a model definite enough to
be tested (plasma universe).  Until these models (or another
model) can account for the cosmological data that have been firmly
established (expansion, CBR, light elements, and structure formation),
the standard cosmology is without a serious competitor.

\subsection{Dark Matters}

An accurate inventory of matter in the Universe still
eludes cosmologists.  What we do know is:  (i) luminous
matter (i.e., matter closely associated with bright stars)
contributes a fraction of the critical density that is
about $0.003h^{-1}$ \cite{fg}; (ii) based upon big-bang nucleosynthesis
baryons contributions a fraction of critical density between $0.009h^{-2}$
and $0.022h^{-2}$ \cite{copi}, which for a generous range of the
Hubble constant corresponds to between about 0.01 and 0.15
of the critical density; (iii) there are indications that the
fraction of critical density contributed by all forms
of matter is {\it at least} $0.1-0.3$ \cite{trimble}---flat rotation curves
of spiral galaxies, virial mass determinations of rich
clusters---and perhaps around the critical density---the peculiar
motions of galaxies, cluster mass determinations based upon
gravitational lensing and x-ray measurements \cite{trimble,dekel}.
(Here the Hubble constant $H_0 = 100h\kms \Mpc^{-1}$ and
the critical density $\rho_{\rm crit} = 3H_0^2/8\pi G =
1.88h^2\times 10^{-29}\gcmm3 \simeq 1.05h^2\times 10^4\eV\cmm3$.)

{}From this one concludes that:  (i) most of the matter in
the Universe is dark; (ii) most of the baryons
are dark; (iii) the dark matter is not closely associated
with bright stars, i.e., it is more diffusely distributed,
e.g., in the extended halos of spiral galaxies; and (iv)
if the total mass density is greater than about 20\% of the
critical density, then there must be another form of matter
since baryons can at most account for 15\% of the critical
density (and only for a low value of the Hubble constant).

The case for $\Omega_0\ga 0.2$ and nonbaryonic dark matter
receives additional support, albeit indirectly,
from other lines of reasoning.  First, it is difficult
to reconcile all the data concerning the formation of
structure in the Universe with a theory that has no nonbaryonic
dark matter (the one model that {\it may} be able to do so is
Peebles' primeval baryon isocurvature model or PBI \cite{pbi}).
Second, the most compelling and comprehensive
theory of the early Universe, inflation \cite{guth}, predicts a
flat Universe (total energy density equal to the critical density)
and thus requires something other than baryons.  Third,
since the deviation of $\Omega$ from unity grows with
time, if $\Omega_0$ is not equal to unity, the epoch when
$\Omega_0$ just begins to deviate significantly from one is
a special epoch and is today(!) (this is often called the
Dicke-Peebles timing argument).

Last but not least, there are three compelling candidates for
the nonbaryonic dark matter:  an axion of mass between $10^{-6}\eV$
and $10^{-4}\eV$; a neutralino of mass between $10\GeV$ and $1000\GeV$; and
a light neutrino species of mass between $10\eV$ and about
$50\eV$ \cite{procnas}.  By compelling, I mean these
particles arose out of efforts to unify
the forces of Nature, and the fact that a particle was predicted
whose relic mass density is close to critical
is a bonus.  This may be the ``Grand Hint'' or the ``Great Misdirection.''

For the axion, the underlying particle physics is Peccei-Quinn symmetry
which is the most attractive solution to the so-called strong-CP problem
(the fact that standard model of particle physics predicts
the electric dipole moment of the neutron to be almost ten
orders of magnitude larger than the current upper limit).  For
the neutralino, it is supersymmetry, the symmetry that relates
fermions and bosons and which helps to explain the large discrepancy
between the weak scale ($300\GeV$) and the Planck scale and may
hold the key to unifying gravity with the other forces.  Unlike the
axion or the neutralino, neutrinos
are known to exist, come in three varieties, and
have a relic abundance known to three significant figures ($
113\cmm3$ per species); the only issue is their mass.
Almost all attempts to unify the
forces and particles of nature lead to the prediction that neutrinos
have mass, often in the ``eV range'' (meaning anywhere from
$10^{-6}\eV$ or smaller to $\keV$).

The axion and neutralino are referred to as ``cold dark matter''
because they move very slowly (neutralinos because they are
heavy and axions because they were produced in the early Universe
with very small momenta).  Neutrinos on the other hand are referred
to as ``hot dark matter'' because they move rapidly (due to
their small mass).  The distinction between the two is crucial
for structure formation:  at early times neutrinos can ``run out''
of overdense regions and into underdense regions,
damping density perturbations on scales smaller than
those corresponding to superclusters.  This means that in the
absence of additional seed perturbations that don't involve
neutrinos (e.g., cosmic string) the sequence of structure
formation in a hot dark matter universe proceeds from the
``top down:''  objects like superclusters form first
and then fragment into smaller objects (galaxies and the like).
Because there is now much evidence that ``small objects''
(galaxies, quasars, neutral hydrogen clouds, and clusters)
were ubiquitous at redshifts from 1 to 4, and ``large objects''
are just forming today, hot dark matter (without additional
perturbations) is not viable.

To end on a sober note, at present
the data can neither prove nor disprove either:
(i) $\Omega_0 = \Omega_B \simeq 0.15$;
(ii) $\Omega_0 = 1$ with $\Omega_B \sim 0.05$ and
$\Omega_{\rm CDM} \sim 0.95$.  (In the first case the Hubble
constant must be near its lower extreme since the nucleosynthesis
measurement of $\Omega_B = 0.009h^{-2} - 0.022h^{-2}$.)
In any case, I will devote the rest of this paper to the second,
more radical possibility.

\subsection{Coherent picture of structure formation}

Because the energy densities of matter (baryons + CDM?)
and radiation (photons, light neutrinos,
and at early times all the other particles in the thermal plasma)
evolve differently, $R^{-3}$ for matter and $R^{-4}$ for
radiation, the energy density
in radiation exceeded that in matter earlier at early times,
$t\la t_{\rm EQ}\sim 10^4\yr$ ($T\ga T_{\rm EQ} \sim 5\eV$ and $R\la
R_{\rm EQ} \sim 3\times 10^{-5}R_{\rm today}$).
Moreover, matter density perturbations do not grow during
the radiation-dominated era, and thus the formation of
structure did not begin in earnest until the epoch of
matter-radiation equality.  After that, (linear)
perturbations in the matter grow as the scale factor,
for a total (linear) growth factor of around 30,000.
This factor sets the characteristic amplitude of
density perturbations, about $few\times
10^{-5}$ (nonlinear structures have formed by the present)
and thus the expected size of temperature fluctuations in the CBR
(density perturbations lead to comparable sized fluctuations
in the CBR temperature).

The detection of CBR anisotropy at the level of about $10^{-5}$
validates the gravitational instability picture of structure
formation.  This success should be viewed in the same way
that the evidence for a large primeval mass
fraction of $^4$He validated the the basic idea of
primordial nucleosynthesis in the late 1960s.
{}From this early success, big-bang nucleosynthesis developed
into a coherent and detailed
explanation for the abundances of D, $^3$He, $^4$He and $^7$Li,
and now provides the earliest test of the big bang,
the most reliable determination of the baryon density,
and an important probe of particle physics.  It is not unreasonable
to hope that a detailed and coherent picture of structure
formation develop and will lead to similar advances in our
understanding of the Universe.

The two crucial elements must underlay any detailed picture:
specification of the quantity and composition of the dark matter and
the nature of the density perturbations.  With regard to the
latter, what is wanted is a mathematical description of the
spectrum of density perturbations.  For example, the Fourier
components $\delta_k$ of the density field and their statistical properties.

There are now three viable theories:  cold dark matter
models; topological-defect models \cite{topo}; and
the primeval baryon isocurvature
model (PBI) \cite{pbi}.  The effort being brought to
bear on this problem---both
experimental and theoretical---is great, and I am confident that
{\it at least} two of these models, if not all three (!), will be
falsified soon.  It is my view that only cold dark matter will survive
the next cut, but of course others may hold a different opinion.

Topological defect models, where
the seeds are cosmic string, monopoles or textures produced
in an early Universe phase transition and the dark matter
is either neutrinos (cosmic string) or cold dark matter (textures),
seem to predict CBR anisotropy on the degree scale that is significantly
less than that measured.  In addition, when normalized to
the COBE measurements of anisotropy, they require a high
level of ``bias;'' bias refers to the discrepancy between the light
and mass distributions, $b \simeq (\delta n_{\rm GAL}/n_{\rm GAL})/
(\delta \rho /\rho )$, which is generally believe to be
of order $1-2$.  Much of the difficulty in assessing the
defect models is on the theoretical side; density perturbations
are constantly being produced as the defect network evolves
and thus cannot easily be described by Fourier components whose
evolution is simple.

The basic philosophy behind the PBI model is
to explain the formation of structure by using ``what is here,''
rather then what early-Universe theorists (like myself) hope is here!
The parameters for PBI are:  $\Omega_0 = \Omega_B\sim 0.2$ and
$H_0 \sim 70\kms\Mpc^{-1}$.  An arbitrary power-law
spectrum of fluctuations in the local baryon number (cut off
at small scales to avoid difficulties with primordial nucleosynthesis)
is postulated and its parameters (slope and normalization)
are determined by the data (CBR fluctuations and
large-scale structure).   PBI has some serious problems:
the baryon density violates the nucleosynthesis bound by
a wide margin ($\Omega_B h^2 \sim 0.1 \gg 0.02$;
it is difficult to make PBI consistent with the
measurements of CBR anisotropy \cite{husugi}.  To wit, Peebles has considered
variations on the basic theme \cite{pbi2} (e.g., adding a cosmological
constant, or even cold dark matter).  At the very least PBI provides
a useful model against which scenarios that postulate exotic dark matter
can be compared; at best, it may represent our Universe.

\section{Inflation and Cold Dark Matter}

Inflation represents a bold attempt to extend
the standard big-bang cosmology to times as early as $10^{-32}\sec$
and to resolve some of the most fundamental questions in
cosmology.  In particular, inflation addresses squarely both the dark
matter and structure formation problems, as well as providing
an explanation for the flatness and smoothness of the Universe.
If successful, inflation would be a truly remarkable addition to
the standard cosmology.

At present there is no standard model of inflation; however, there are
many viable models, all based on well defined speculations about
physics at energy scales of around $10^{14}\GeV$ and higher \cite{inflate}.
Inflation makes three robust predictions:  (1) spatially flat
Universe \cite{differ}; (2) nearly scale-invariant spectrum of
density (scalar metric)
perturbations \cite{scalar}; (3) nearly scale-invariant spectrum
of gravity waves (tensor metric perturbations) \cite{tensor}.

With regard to metric perturbations; they are imprinted during
inflation, arising from quantum mechanical fluctuations
excited on extremely small scales ($\la 10^{-23}\rcm$),
which are stretched to astrophysical
scales ($\ga 10^{25}\rcm$) by the tremendous growth
in the scale factor during inflation ($\ga 10^{25}$).
In almost all models of inflation
the statistics of the perturbations are gaussian,
and the Fourier power spectrum, $P(k) \equiv |\delta_k|^2$,
completely specifies the statistical properties of the density field.

While the metric perturbations are predicted to nearly
scale invariant, the small deviations that can occur encode
much about the underlying inflationary model.  Likewise, the
amplitudes of the metric perturbations are model dependent and
hold equally important information.  (Scale-invariant
density perturbations means fluctuations in the gravitational
potential that are equal on all scales at early times;
for the gravitational waves, scale invariant means that all
gravity waves cross the horizon with equal amplitude.)

The first prediction means the total energy density
(including matter, radiation, and
the vacuum energy density associated with a cosmological constant)
is equal to the critical density, that is $\Omega_0 =1$.
Coupled with our knowledge of the baryon density, this implies that
the bulk of matter in the Universe (95\% or so) must be nonbaryonic.  The
two simplest possibilities are hot dark matter or cold dark matter.
Structure formation with hot dark matter has been studied,
and, sadly, does not work; thus we are led to to cold dark matter.

For cold dark matter there is no damping of perturbations
on small scales, and structure is built from the ``bottom up:''
Clumps of dark matter and baryons continuously merge to form larger
objects.  ``Typical galaxies'' are formed at redshifts
$z\sim 1-2$; ``rare objects'' such as quasars and radio
galaxies can form earlier from regions where the density
perturbations have larger than average amplitude.  Clusters form
in the very recent past (redshifts less than order unity), and
superclusters are just forming today.  Voids naturally arise as regions
of space are evacuated to form objects \cite{primack}.

\subsection{Almost, but is something missing?}

Broadly speaking, testing the cold dark matter scenario involves
measuring the quantity, composition, and distribution
of dark matter and determining the spectrum of density
perturbations.   I have already discussed the current state of
our knowledge of dark matter.   While a host of observations
provide information about the primeval spectrum of density
perturbations, measurements of the anisotropy of the CBR and
mapping the distribution of matter today (as traced
by bright galaxies) are perhaps most crucial.  (For reference,
perturbations on scales of about $1\Mpc$ correspond to galactic
sized perturbations, on $10\Mpc$ to cluster size perturbations,
on $30\Mpc$ to the large voids, and $100\Mpc$ to the great walls.)

CBR anisotropy probes the power spectrum on large scales.  The
CBR temperature difference measured on a given angular scale
is related to the power spectrum on a given length scale:
$\lambda \sim (\theta /{\rm deg})100h^{-1}\Mpc$.
Since the COBE detection, a host of ground-based and balloon-borne
experiments have also detected CBR anisotropy, on scales from
about $0.5^\circ$ to $90^\circ$, at the level of around $30\mu$K ($\delta
T/T \sim 10^{-5}$).  The measurements are consistent with
the predictions of cold dark matter, though there are still
large statistical uncertainties as well as concerns about
contamination by foreground sources \cite{wss}.  There is
a great deal of experimental activity (more than ten groups),
and measurements in the
near future should improve the present situation significantly.
The CBR contains important information on angular scales down to
about $0.1\,$deg (anisotropy on smaller angular scales is washed
out due to the finite thickness of the last scattering surface).
A follow-on to COBE, being studied in both Europe and the US,
and a variety of earth-based and balloon-based
experiments should hopefully map CBR anisotropy on scales from
$0.1\,$deg to $90\,$deg in the next decade.

The COBE detection of CBR anisotropy not only provided the
first evidence for the existence of primeval density perturbations,
but also an unambiguous way to normalize
the spectrum of density perturbations:  Given the shape of
the power spectrum (for cold dark matter, approximately scale
invariant) the COBE measurement (on a scale of around $10^3h^{-1}\Mpc$)
ties down the spectrum on all scales.  This
leads to definite predictions that can be tested by other CBR
measurements and observations of large-scale structure.

The comparison of predictions for structure formation
with present-day observations of the
distribution of galaxies is very important, but fraught with difficulties.
Theory most accurately predicts ``where the mass is''
(in a statistical sense) and the observations determine where the light is.
Redshift surveys probe present-day inhomogeneity on scales
from around one $\Mpc$ to a few hundred $\Mpc$, scales where
the Universe is nonlinear ($\delta n_{\rm GAL}/n_{\rm GAL}
\ga 1$ on scales $\la 8h^{-1}\Mpc$) and where astrophysical
processes undoubtedly play an important role
(e.g., star formation determines where and when
``mass lights up,'' the explosive release of energy in supernovae
can move matter around and influence subsequent star formation,
and so on).  The distance to a galaxy is
determined through Hubble's law ($d = H_0^{-1} z$) by measuring a redshift;
peculiar velocities induced by the lumpy distribution
of matter are significant and prevent a direct determination
of the actual distance.  There are the intrinsic limitations
of the surveys themselves:  they are flux not volume limited (brighter
objects are seen to greater distances and vice versa) and relatively
small (e.g., the CfA slices of the Universe survey contains only
about $10^4$ galaxies and extends to a redshift of about $z\sim 0.03$).
Last but not least are the numerical
simulations which bridge theory and observation;
they are limited dynamical range (about a factor
of 100 in length scale) and in microphysics (in the largest simulations
only gravity, and in others only a gross approximation to the effects of
hydrodynamics/thermodynamics).

This being said, redshift surveys do provide an important probe of
the power spectrum on small scales ($\lambda \sim 1 - 300
\Mpc$).  Even with their limitations redshift surveys (as well as other
data) indicate that while
the simplest version of COBE-normalized cold dark matter
is in broad agreement with the data, the shape
of the power spectrum as well as its amplitude
on small scales is not quite right \cite{jpo,ll}.  At least three
possibilities come to mind:  (i) the comparison
of numerical simulations and the
observations is still too primitive to draw firm conclusions;
(ii) cold dark matter has much, but not all, of the ``truth;''
or (iii) cold dark matter has been falsified.

For three reasons I believe that it is worthwhile
exploring possibility (ii), namely
that something needs to be added to cold dark matter.
First, cold dark matter is such an attractive theory and part
of a bold attempt to extend greatly the standard cosmology.
Second, many observations seem to point to
the same problem (e.g., the abundance of x-ray clusters
and the cluster-cluster correlation function).  Third,
there are other reasons to believe that the Universe is more
complicated than the simplest model of cold dark matter.

\subsection{Five cold dark matter models}

Somewhat arbitrarily, standard cold dark matter has come to mean:
precisely scale-invariant
density perturbations; baryons + CDM only; and Hubble constant of
$50\kms\Mpc^{-1}$ (to ensure a sufficiently aged Universe with a
Hubble constant still within the range of observations).  This
is the vanilla or default model, which, when normalized to COBE has
too much power on small scales and the wrong spectral shape on larger scales.

The spectrum of density perturbations today depends
not only upon the primeval spectrum (and the normalization
on large scales provided by COBE), but also upon the energy content
of the Universe.  While the fluctuations in the gravitational potential
were initially approximately scale invariant, the fact that
the Universe evolved from an early radiation-dominated phase
to a matter-dominated
phase imposes a characteristic scale on the spectrum of density
perturbations seen today; that scale is determined by the energy
content of the Universe (the characteristic scale
$\lambda_{\rm EQ}\sim 10h^{-1}\Mpc g_*^{1/2}/\Omega_{\rm matter}h$
where $g_*$ counts the relativistic degrees of freedom and
$\Omega_{\rm matter} = \Omega_B +\Omega_{\rm CDM}$).
In addition, if some of the nonbaryonic dark
matter is neutrinos, they will inevitably suppress power on small
scales through freestreaming.  With this in mind, let me discuss
the small modifications of cold dark matter that improve its agreement
with the observations.

(1) Low Hubble Constant + cold dark matter (LHC CDM).  Remarkably, simply
lowering the Hubble constant to around $30\kms\Mpc^{-1}$ solves all the
problems
of cold dark matter.  Recall, the critical density $\rho_{\rm crit}
\propto H_0^2$; lowering $H_0$ lowers the matter density and postpones
matter-radiation equality, which has precisely the desired effect
on the spectrum of perturbations.  It has two other added benefits:
it makes the expansion age of the Universe comfortably
consistent with the ages of the
oldest stars and raises the baryon fraction of critical
density to a value that is consistent with that measured in x-ray
clusters (see below).  Needless to say, such a
small value for the Hubble constant flies in the face of current
observations; further, it illustrates the fact that the problems of
cold dark matter get even worse for the larger values of $H_0$
that have been determined by recent observations \cite{hubble}.

(2) Hot + cold dark matter ($\nu$CDM).  Adding a small amount of
hot dark matter can suppress density perturbations on small scales;
of course, too much leads back to the longstanding
problems of hot dark matter.  The amount required is
about 20\%, corresponding to about ``$5\eV$ worth
of neutrinos'' (i.e., one species of mass $5\eV$, or two species
of mass $2.5\eV$, and so on).  This admixture of hot dark matter
rejuvenates cold dark matter provided the Hubble constant is not
too large, $H_0\la 55 \kms\Mpc^{-1}$.

(3) Cosmological constant + cold dark matter ($\Lambda$CDM).
(A cosmological constant corresponds to a uniform energy density,
or vacuum energy.)  Shifting 60\% to 80\% of the critical density
to a cosmological constant lowers the matter density
and has the same beneficial effect as a low Hubble constant.
In fact, a Hubble constant as large as $80\kms\Mpc^{-1}$
can be tolerated.  In addition,
the cosmological constant allows the age problem to solved
even if the Hubble constant is large, addresses the fact
that few measurements of the mean mass density give a value as large as
the critical density (most measurements of the mass density
are insensitive to a uniform component), and allows the
fraction of matter in baryons to be large (see below).
Not everything is rosy;
cosmologists have invoked a cosmological constant twice before to solve
their problems (Einstein to obtain a static universe and
Bondi, Gold, and Hoyle to solve the earlier age crisis when
$H_0$ was thought to be $250\kms\Mpc^{-1}$).  Further, particle
physicists can still not explain why the energy of the
vacuum is not at least 50 (if not 120) orders of magnitude larger than
the present critical density.

(4)  Extra relativistic particles + cold dark matter ($\tau$CDM).
The epoch of matter-radiation equality can also be delayed by
raising the level of radiation.  In the standard cosmology the
radiation content today consists of photons + three (undetected)
cosmic seas of neutrinos (corresponding to $g_* \simeq
3.36$).   While we have no direct determination
of the radiation beyond that in the CBR, there are
at least two problems:  (i) what are the additional relativistic
particles?; and (ii) can additional radiation be added without
upsetting the successful predictions
of primordial nucleosynthesis which depend critically upon the
energy density of relativistic particles.  The simplest way around
these problems is an unstable tau neutrino (mass anywhere
between a few keV and a few MeV) whose decays produce
the radiation.   This fix can tolerate a larger Hubble constant,
though at the expense of more radiation.

(5)  Tilted cold dark matter (TCDM).  While the spectrum of density
perturbations in most models of inflation is very nearly scale invariant,
there are models where the deviations are significant and lead to
smaller fluctuations on small scales.  Further, not
only do density perturbations produce CBR anisotropy, but so do the
gravitational waves; if gravity waves account for a significant
part of the CBR anisotropy, the level of density perturbations must be
lowered.  A combination of tilt and gravity waves can solve the
problem of too much power on small scales, but does not seem to address
the shape problem as well as the other fixes.

In evaluating these better fit models, one should keep the words
of Francis Crick in mind (loosely paraphrased):  A model that fits
all the data at a given time is necessarily wrong, because at any given
time not all the data are correct(!).  $\Lambda$CDM provides an
interesting example; when I discussed it in 1990, I called it the
best-fit Universe, but not the best motivated and was certain it
would fall by the wayside \cite{lcdm}.  In 1995, it is still
probably the best-fit model.

Let me end by defending the other point of view, namely,
that to add something to cold dark matter is not unreasonable,
or even as some have said, a last gasp effort to saving a dying theory.
Standard cold dark matter was a starting point, similar to
early calculations of big-bang nucleosynthesis.  It was always
appreciated that the inflationary spectrum of density perturbations
was not exactly scale invariant \cite{pjs} and that the Hubble constant
was unlikely to be exactly $50\kms \Mpc$.  As the quality and quantity
of data improve, it is only sensible to refine the model, just as has been
done with big-bang nucleosynthesis.  Cold dark matter seems to
embody much of the ``truth.''   The modifications suggested
all seem quite reasonable (as opposed
to contrived).  Neutrinos exist; they are expected to have mass;
there is even some experimental data that indicates they do have
mass.  It is still within the realm of possibility that the Hubble
constant is less than $50\kms\Mpc^{-1}$, and if it is as large
as $70\kms\Mpc^{-1}$ to $80\kms\Mpc^{-1}$ a cosmological constant
seems inescapable based upon the age problem.  There is no data
that can preclude more radiation than in the standard cosmology
and deviations from scale invariance were always expected.

\section{The Future}
\subsection{Testing and discriminating}

The stakes for cosmology are high:  if correct, inflation/cold dark matter
represents a major extension of the big bang and our understanding
of the Universe, which can't help but shed
light on the fundamental physics at energies of order $10^{14}\GeV$ or higher.

How and when we will have definitive tests of cold dark matter?
Because of the large number of measurements that are being carried
out and can have significant impact,
I believe sooner rather than later.  The list is long:
CBR anisotropy; larger redshift surveys (e.g., the Sloan Digital
Sky Survey will have $10^6$ redshifts); direct searches
for the nonbaryonic in our neighborhood (e.g., axion and neutralino searches)
and baryonic dark matter (microlensing); x-ray studies of galaxy
clusters; the use of back-lit gas clouds (quasar absorption line systems)
to study the Universe at high redshift; galactic evolution (as
revealed by deep images of the sky taken by the Hubble Space Telescope and
Keck 10 meter telescope); a variety of measurements of $H_0$ and $q_0$;
mapping of the peculiar velocity field at large redshifts
through the Sunyaev-Zel'dovich effect; dynamical estimates of
the mass density (weak gravitational lensing, large-scale velocity
fields, and so on); age determinations of the Universe;
gravitational lensing; searches for supersymmetric particles (at accelerators)
and neutrino oscillations (at accelerators, solar-neutrino detectors,
and other large underground detectors); searches for high-energy neutrinos from
neutralino annihilates in the sun using large underground detectors;
and on and on.  Consider the possible impact of a few specific examples.

A definitive determination that $H_0$ is greater than $55\kms\Mpc^{-1}$ would
falsify LHC CDM and
$\nu$CDM.  Likewise, a definitive determination that $H_0$ is $75\kms
\Mpc^{-1}$ or larger would necessitate a cosmological constant.  A flat
Universe
with a cosmological constant has a very different deceleration
parameter than one dominated by matter, $q_0 =-1.5\Omega_\Lambda + 0.5
\sim -(0.4 - 0.7)$ compared to $q_0 = 0.5$, and this could be settled
by galaxy number counts or numbers of lensed quasars.  The level
of CBR anisotropy in $\tau$CDM and LHC CDM on the $0.5^\circ$ scale
is about 50\% larger than the other models, which should be
easily measurable.   If neutrino-oscillation
experiments were to provide evidence for a neutrino of mass $5\eV$ (two of
mass $2.5\eV$) $\nu$CDM would seem almost inescapable.

A map of the CBR with $0.5^\circ -1^\circ$ resolution could separate the
gravity-wave
from density perturbation contribution to the CBR anisotropy and
provide evidence for the third robust prediction of inflation.  Further,
mapping CBR anisotropy on these scales or slightly smaller offers
the possibility determining the geometry of the Universe (the position
of the ``Doppler'' peak scales as $0.5^\circ /\sqrt{\Omega_0}$ \cite{omega}).

X-ray observations of rich clusters are able to determine the ratio
of hot gas (baryons) to total cluster mass (baryons + CDM) (by a wide
margin, most of the baryons ``seen'' in cluster are in the hot gas).
To be sure there are assumptions and uncertainties; the data at the moment
indicate that this ratio is $0.07h^{-3/2}$ \cite{henry}.  If clusters provide
a fair sample of the universal mix of matter, then this ratio should
equal $\Omega_B/(\Omega_B + \Omega_{\rm CDM}) \simeq (0.009-0.022)h^{-2}/
(\Omega_B + \Omega_{\rm CDM})$.  Since clusters are large objects
they should provide an approximately fair sample.  Taking the
numbers at face value, cold dark matter is consistent with the
cluster gas fraction provided either:  $\Omega_B + \Omega_{\rm CDM}
= 1$ and $h\sim 0.3$ or $\Omega_B + \Omega_{\rm CDM} \sim 0.3$ and
$h\sim 0.7$, favoring LHC CDM or $\Lambda$CDM.

If cold dark matter is correct, then a significant,
if not dominant, fraction of the dark halo of our galaxy should be cold
dark matter (the halos of spiral galaxies are not large enough to
guarantee that they represent a fair sample).  Direct searches for
faint stars have failed to turn up enough to account for
the halo \cite{bahcall}.  Over the past few years, microlensing has been
used to search for dark stars (stars
below the $0.08M_\odot$ limit for hydrogen burning).  Five stars
in the LMC have been observed to change brightness in a way consistent
with their being microlensed by dark halo objects passing
along the line of sight.  While the statistics are
small, and there are uncertainties concerning the size of dark
halo, these results indicate that only a small fraction (5\% to
30\%) of the dark halo is in the form of dark stars \cite{fewmachos}.

\subsection{Reconstruction}

If cold dark matter is shown to be correct, then a window to the
very early Universe ($t\sim 10^{-34}\sec$) will
have been opened.  While it is certainly premature to jump to this
conclusion, I would like to illustrate one example of what one could
hope to learn.  As mentioned earlier, the spectra and amplitudes
of the the tensor and scalar metric perturbations predicted by
inflation depend upon the underlying model, to be specific, the
shape of the inflationary scalar-field potential.
(Inflation involves the classical evolution of a scalar
field $\phi$ rolling down its potential
energy curve $V(\phi )$.)  If one can measure the power-law
index of the scalar spectrum and the amplitudes of the scalar
and tensor spectra, one can recover the value of the potential
and its first two derivatives around the point on the potential
where inflation took place \cite{recon}.
(Measuring the power-law index of the
tensor perturbations in addition, allows an important
consistency check of inflation.)   Reconstruction
of the inflationary scalar potential would shed light both on inflation
as well as physics at energies of the order of $10^{14}\GeV$.

\subsection{Concluding remarks}

We live in exciting times.  We have a cosmological model that
provides a reliable accounting of the Universe from 0.01 sec
until the present.  Together with the standard model of particle
physics it provides a framework for both asking deeper
questions about the Universe and making sensible speculations.  With
inflation and cold dark matter we may be on the verge of a very
significant extension of the standard cosmology.  Most importantly,
the data needed to test the cold dark matter theory is coming
in at a rapid rate.  At the very least we should soon know whether
we are on the right track or if it's back to the drawing board.

\bigskip\bigskip\bigskip
This work was supported in part by the DOE (at Chicago and Fermilab)
and by the NASA through grant NAG 5-2788 (at Fermilab).

\end{document}